\documentclass[english,12pt]{article}
          \usepackage{babel}
          \usepackage[T1]{fontenc}
          \usepackage[latin2]{inputenc}
          \usepackage{pslatex}
\begin{document}

\title{\bf Curious Variables Experiment (CURVE). \\
TT Bootis - superhump period change pattern confirmed.}
\author{A. ~O~l~e~c~h$^1$, ~L.M. ~C~o~o~k$^2$, ~K. ~Z~{\l}~o~c~z~e~w~s~k~i$^3$, 
~K. ~M~u~l~a~r~c~z~y~k$^3$, \\ ~P. ~K~\c{e}~d~z~i~e~r~s~k~i$^3$, ~A. 
~U~d~a~l~s~k~i$^3$ ~and~ ~M. ~W~i~{\'s}~n~i~e~w~s~k~i$^1$}
\date{$^1$ Nicolaus Copernicus Astronomical Center,
Polish Academy of Sciences,
ul.~Bartycka~18, 00-716~Warszawa, Poland\\
{\tt e-mail: (olech,mwisniew)@camk.edu.pl}\\
~\\
$^2$ Center for Backyard Astrophysics (Concord), \\ 1730 Helix Court, Concord,
CA 94518, USA \\ {\tt e-mail: lcoo@yahoo.com}\\ 
~\\
$^3$ Warsaw University Observatory, Al. Ujazdowskie 4, 00-476 Warszawa,
Poland\\ {\tt e-mail: (kzlocz,kmularcz,pkedzier,udalski)@astrouw.edu.pl}}
\maketitle

\begin{abstract} 

We report extensive multi-station photometry of TT Boo during its June
2004 superoutburst. The amplitude of the superoutburst was about 5.5 mag
and its length over 22 days. The star showed a small re-brightening
starting around the 9th day of the superoutburst. During entire bright
state we observed clear superhumps with amplitudes from 0.07 to 0.26 mag
and a mean period of $P_{sh} = 0.0779589(47)$ days ($112.261\pm0.007$
min). The period was not constant but decreased at the beginning and
the end of superoutburst yet increased in the middle phase. We argue
that the complicated shape of the $O-C$ diagram is caused by real
period changes rather than by phase shifts. Combining the data from two
superoutbursts from 1989 and 2004 allowed us to trace the birth of the
late superhumps and we conclude that it is a rather quick process
lasting about one day.

\noindent {\bf Key words:} Stars: individual: TT Boo -- binaries:
close -- novae, cataclysmic variables
\end{abstract}

\section{Introduction}

$UBV$ photometry of TT Boo in quiescence was obtained by Szkody (1987).
She found values of $B-V=0.38$ and $U-B=-1.09$ mag, which are rather
typical for dwarf novae of comparatively long periods but not for SU
UMa stars which usually have $B-V$ color around zero. Howell \& Szkody
(1988) reported quiescent photometry of TT Boo and their observations
revealed light variations with a period near $111\pm5$ min with an
amplitude of 0.2 mag. They also mention the observations of Thorstensen
and Brownsberger, who observed the star in the bright state and found
features that looked like superhumps with a tentative period of 97 min.
This is different from observations of Howell \& Szkody (1988) because
in all confirmed typical SU UMa stars the superhump period is slightly
longer than the orbital period. Spectroscopy of TT Boo in the outburst
was obtained by Bruch (1989). 

More detailed observations of TT Boo in its bright state were performed
during two nights of April 1993 by Kato (1995). He found clear
superhumps with a period of $0.07811(5)$ days confirming that TT Boo
belongs to the SU UMa-type dwarf novae class. Despite its quite frequent
outbursts and brightness at maximum of $mV = 12.7$ mag, TT Boo is a
poorly studied object. The best determination of the superhump period
is based on only two nights of observations and the orbital period of
the system is not known. 

We were alerted to the ongoing outburst of TT Boo by Carlo Gualdoni's
VSNET outburst alert number 6316. He reported that on June 3.9792 UT
the star was at magnitude 12.8.

\section{Observations and Data Reduction}

Observations of TT Boo reported in the present paper were obtained
during two superoutbursts. Observations from August 1989 were collected
at Dominion Astrophysical Observatory (DAO) Victoria B.C., Canada with
the 1.22-m telescope equipped with an RCA-2 CCD camera. A Johnson $V$
filter was used. The exposure times were 60 and 120 seconds on the first
and second nights, respectively. 

The data from 2004 were collected at two locations: the Ostrowik station
of the Warsaw University Observatory and CBA Concord in the San
Francisco suburb of Concord, approximately 50 km from East of the City.
The Ostrowik data were collected using the 60-cm Cassegrain telescope
equipped with a Tektronics TK512CB back-illuminated CCD camera. The
scale of the camera was 0.76"/pixel providing a 6.5' x 6.5' field of
view. The full description of the telescope and camera was given by
Udalski and Pych (1992). The CBA data were collected using an f/4.5 73-
cm reflector operated at prime focus on an English cradle mount. Images
were collected with a Genesis G16 camera using a KAF1602e chip giving a
field of view of 14.3' x 9.5'. Images were reduced using AIP4WIN
software (Berry and Burnell 2000).

In Ostrowik and CBA Concord the star was monitored in "white light" in
order to be able to observe it also at minimum light of around 19 mag.
We used two comparison stars: GSC 3047:313 ($RA = 14^h57^m55.03^s$,
Decl.$ = +40^\circ45' 17"$) and GSC 3047:41 ($RA = 14^h57^m43.0^s$,
Decl.$ = +40\circ45'06"$). CBA Concord exposure times were 15, 20 and 30
seconds depending upon the brightness of the star. The Ostrowik exposure
times were from 90 to 150 seconds during the bright state and from 150
to 240 seconds at minimum light. A full journal of our CCD observations
of TT Boo is given in Table 1. In 2004, we monitored the star for 69
hours during 25 nights and obtained 3924 exposures. In 1989, during two
nights, we collected 366 exposures and followed the star for a total
time of 8.47 hours. 

\begin{table}[!h]
\caption{\sc Journal of the CCD observations of TT Boo.}
\vspace{0.1cm}
\begin{center}
{\small
\begin{tabular}{|l|c|c|c|r|l|}
\hline
\hline
Date & No. of & Start & End & Length & Location \\
     & frames & 2447000. + & 2447000. + & [hr]~ & \\
\hline
1989 Aug 04/05 &  252   & 742.70840 & 742.89724 & 4.532 & DAO \\
1989 Aug 05/06 &  114   & 743.70928 & 743.87345 & 3.940 & DAO \\
\hline
Total          &  366   & -- & -- & 8.472 & \\
\hline
\hline
Date & No. of & Start & End & Length & Location \\
     & frames & 2453000. + & 2453000. + & [hr]~ & \\
\hline
2004 Jun 04/05 &  564   & 161.69325 & 161.94452 &  6.030  & CBA Concord \\ 
2004 Jun 05/06 &  564   & 162.70358 & 162.95793 &  6.104  & CBA Concord \\
2004 Jun 06/07 &   55   & 163.44986 & 163.54116 &  2.191  & Ostrowik\\
2004 Jun 07/08 &   48   & 164.46096 & 164.53105 &  1.682  & Ostrowik\\
2004 Jun 08/09 &   73   & 165.37321 & 165.52394 &  3.618  & Ostrowik\\
2004 Jun 09/10 &  128   & 166.35531 & 166.53572 &  4.330  & Ostrowik\\
2004 Jun 10/11 &   61   & 167.43234 & 167.51018 &  1.868  & Ostrowik\\
2004 Jun 12/13 &   78   & 169.35002 & 169.50455 &  3.709  & Ostrowik\\
2004 Jun 13/14 &  110   & 170.34641 & 170.53687 &  4.571  & Ostrowik\\
2004 Jun 14/15 &  286   & 171.69416 & 171.91756 &  5.362  & CBA Concord \\
2004 Jun 15/16 &  616   & 172.70006 & 172.94261 &  5.821  & CBA Concord \\
2004 Jun 16/17 &  412   & 173.69051 & 173.89688 &  4.953  & CBA Concord \\
2004 Jun 17/18 &  171   & 174.73610 & 174.82214 &  2.065  & CBA Concord \\
2004 Jun 18/19 &   15   & 175.70148 & 175.81575 &  0.301  & CBA Concord \\
2004 Jun 19/20 &  104   & 176.69015 & 176.77809 &  2.111  & CBA Concord \\
2004 Jun 20/21 &  375   & 177.69177 & 177.92182 &  5.521  & CBA Concord \\
2004 Jun 21/22 &   74   & 178.35158 & 178.53101 &  4.306  & Ostrowik\\
2004 Jun 21/22 &  160   & 178.69378 & 178.79589 &  2.451  & CBA Concord \\
2004 Jun 24/25 &    4   & 181.36495 & 181.38887 &  0.574  & Ostrowik\\
2004 Jun 29/30 &    1   & 186.36816 & 186.37024 &  0.002  & Ostrowik\\
2004 Jun 30/01 &    5   & 187.39702 & 187.40611 &  0.218  & Ostrowik\\
2004 Jul 02/03 &    6   & 189.45532 & 189.46587 &  0.253  & Ostrowik\\
2004 Jul 04/05 &    5   & 191.42160 & 191.44154 &  0.479  & Ostrowik\\
2004 Jul 06/07 &    5   & 193.39290 & 193.40401 &  0.267  & Ostrowik\\
2004 Jul 09/10 &    4   & 196.36391 & 196.37317 &  0.222  & Ostrowik\\
\hline
Total          &  3924  & -- & -- & 69.01 & \\ 
\hline
\hline
\end{tabular}}
\end{center}
\end{table}

All the Ostrowik and DAO data reductions were performed using a standard
procedure based on the IRAF\footnote{IRAF is distributed by the National
Optical Astronomy Observatory, which is operated by the Association of
Universities for Research in Astronomy, Inc., under a cooperative
agreement with the National Science Foundation.} package and profile
photometry was derived using the DAOphotII package (Stetson 1987). The
typical accuracy of our measurements varied between 0.004 and 0.11 mag
depending on the brightness of the object. The median value of the
photometric errors was 0.015 mag

\section{General light curve}

Figure 1 shows the general light curve of TT Boo during our 2004
campaign. The rough transformation to $V$ magnitude was made using the
comparison star GSC 3047:313 ($RA = 14^h57^m55.3^s$, Decl.$ =
+40^\circ45'17"$, $V = 12.862$, $B-V = 0.889$) and assuming that the
$B-V$ color of TT Boo in superoutburst is around zero (Bruch and Engel
1994). We additionally assumed that the sensitivity of our detector in
"white light" roughly corresponds to Cousins R band (Udalski and Pych
1992) and used Caldwell et al. (1993) transformation between $B-V$ and
$V-R$ colors.

CCD observations are marked with dots while the open square corresponds
to the observation of Carlo Gualdoni from June 3/4 reported to VSNET. The
star was caught by him at the very beginning of the superoutburst
because AAVSO observations from June 1/2 found TT Boo below 17.5 mag.
Thus we conclude that the superoutburst started on June 2 or 3.

\vspace{8.9cm}

\includegraphics{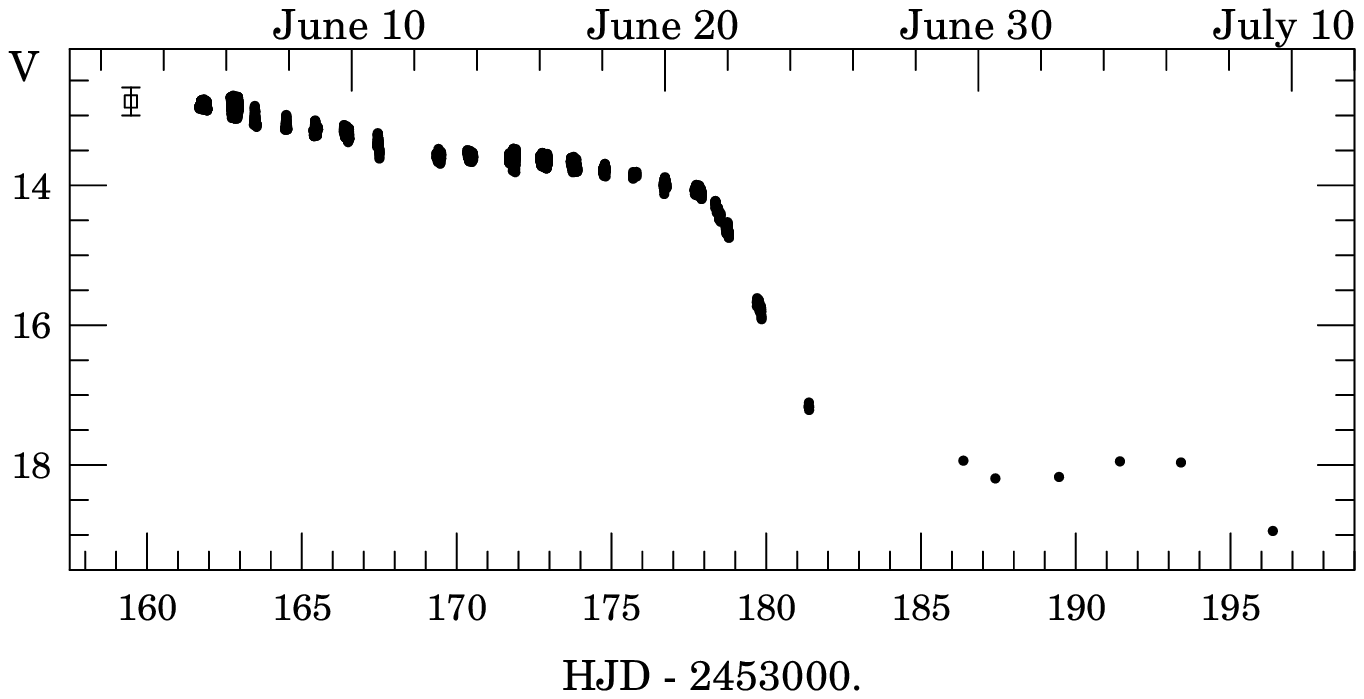}

   \begin{figure}[h]
      \caption{\sf The general photometric behavior of TT Boo during
its 2004 superoutburst. Our CCD observations are marked with dots and
the open square represents the observation of Carlo Gualdoni reported to
VSNET.}
   \end{figure}

Our first observations taken on June 4 between 4:38 and 10:40 UT show a
slight declining trend with slope of 0.04 mag/day. During the next seven
nights the decline was much steeper with a slope of 0.096 mag/day.
Around June 12/13 we noted a clear change of the slope value to 0.073
mag/day. Similar phenomenon is often observed in other SU UMa stars as
was noticed and summarized by Kato et al (2003). Around 12 UT on June
21, TT Boo entered the final decline phase with slope of 1.015 mag/day
reaching magnitude 18 around June 25. The entire superoutburst thus
lasted 22-23 days. From June 22 to July 7 the star stayed at brightness
of 18 mag and on July 9 it finally faded to its quiescent magnitude of
around 19 mag.

\clearpage

~

\vspace{20.7cm}

\includegraphics{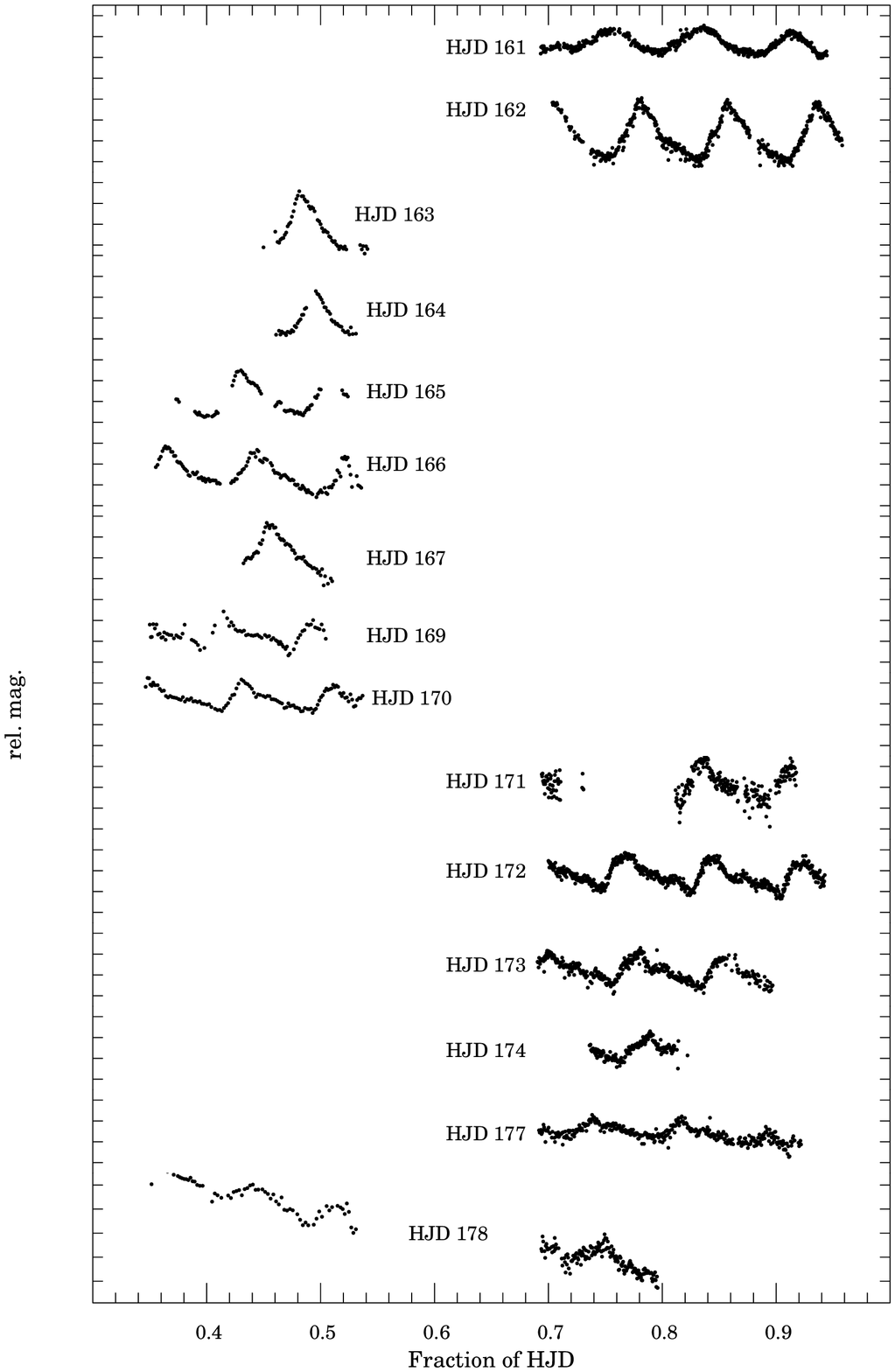}

   \begin{figure}[h]
      \caption{\sf The light curves of TT Boo from its 2004 June
superoutburst. The ticks on the vertical axis are separated by 0.1
mag.}
   \end{figure}

\section{Superhumps}

As shown in Fig. 2, superhumps were present in the light curve of the
dwarf nova on all nights from June 4 till June 21 (HJD from 161 to 178).
It is difficult to recognize them on June 22 when the star entered the
final decline phase. On June 4 the superhumps have an amplitude of only
0.10 mag and a sinusoidal shape. This means that we caught TT Boo very
close to the moment of birth of the superhumps. On June 5 the star
flashed with fully developed characteristic tooth-shape modulations with
amplitude of 0.26 mag. 

The evolution of the amplitude of the superhumps and shape of the light
curve is shown in Fig. 3. This plot shows nightly light curves of TT Boo
phased with the corresponding period (see next sections) and averaged in
0.02-0.05 phase bins. One can clearly see that tooth-shaped and large
amplitude variations were observed from June 5 to 10. Around June 11-12
(HJD 169-170) amplitude significantly decreased and the secondary humps
at phase around 0.3-0.5 became visible.

\vspace{14.6cm}

\includegraphics{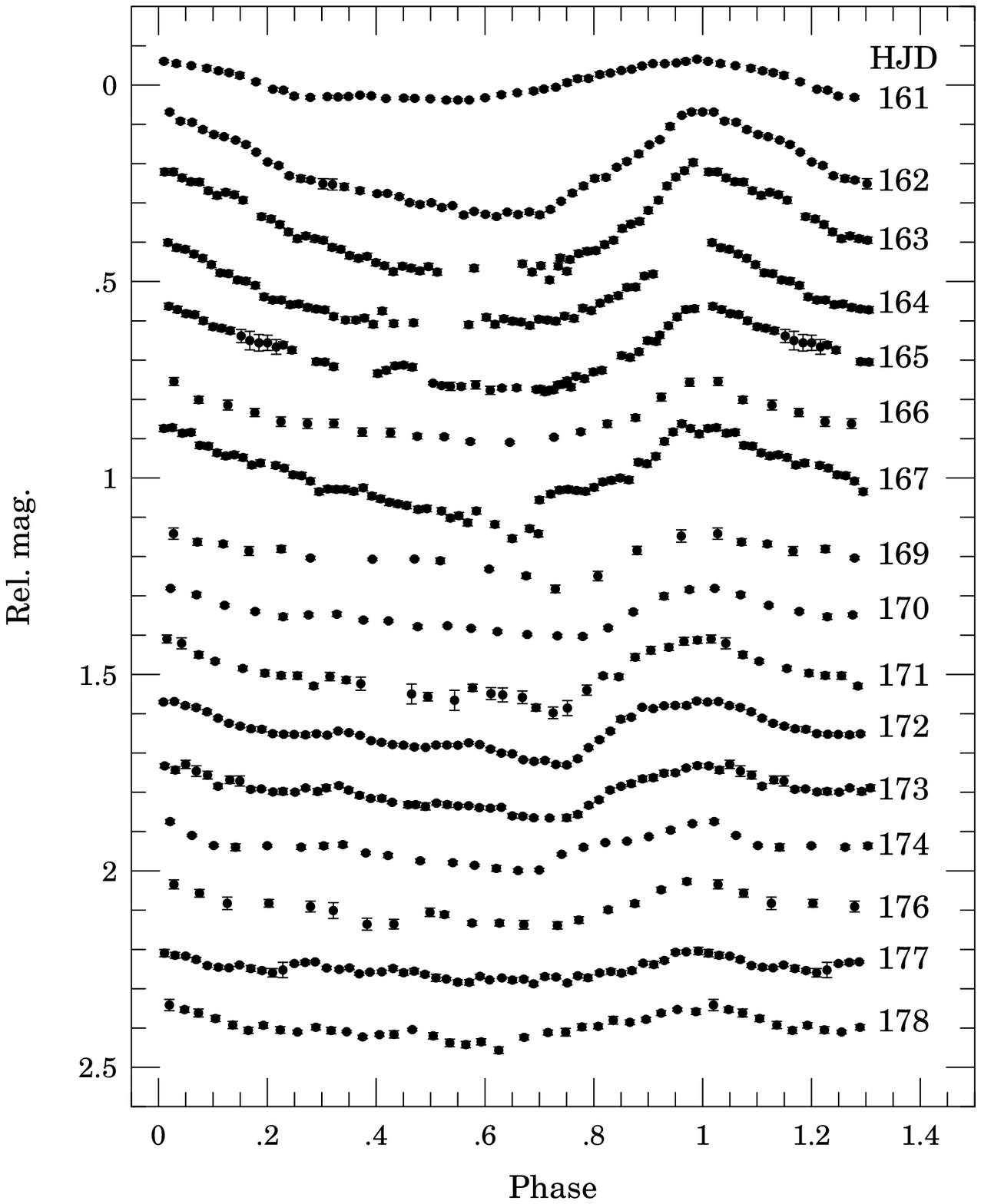}

   \begin{figure}[h]
      \caption{\sf Nightly light curves of TT Boo
phased with the corresponding superhump period and averaged in
0.02--0.05 phase bins.
              }
   \end{figure}

\subsection{Power spectrum of 2004 data}

From each light curve of TT Boo in superoutburst we removed the first or
second order polynomial and analyzed them using {\sc anova} statistics
with two harmonic Fourier series (Schwarzenberg-Czerny 1996). The
resulting periodogram is shown in the upper panel of Fig. 4. The most
prominent peak is found at a frequency of $f_1=12.818\pm0.005$ c/d,
which corresponds to a period of $P_{sh}=0.078015(30)$ days 
($112.34\pm0.04$ min). The first harmonic of this frequency at
$f_2=25.74\pm0.03$ c/d is also clearly visible.

\vspace{13.7cm}

\includegraphics{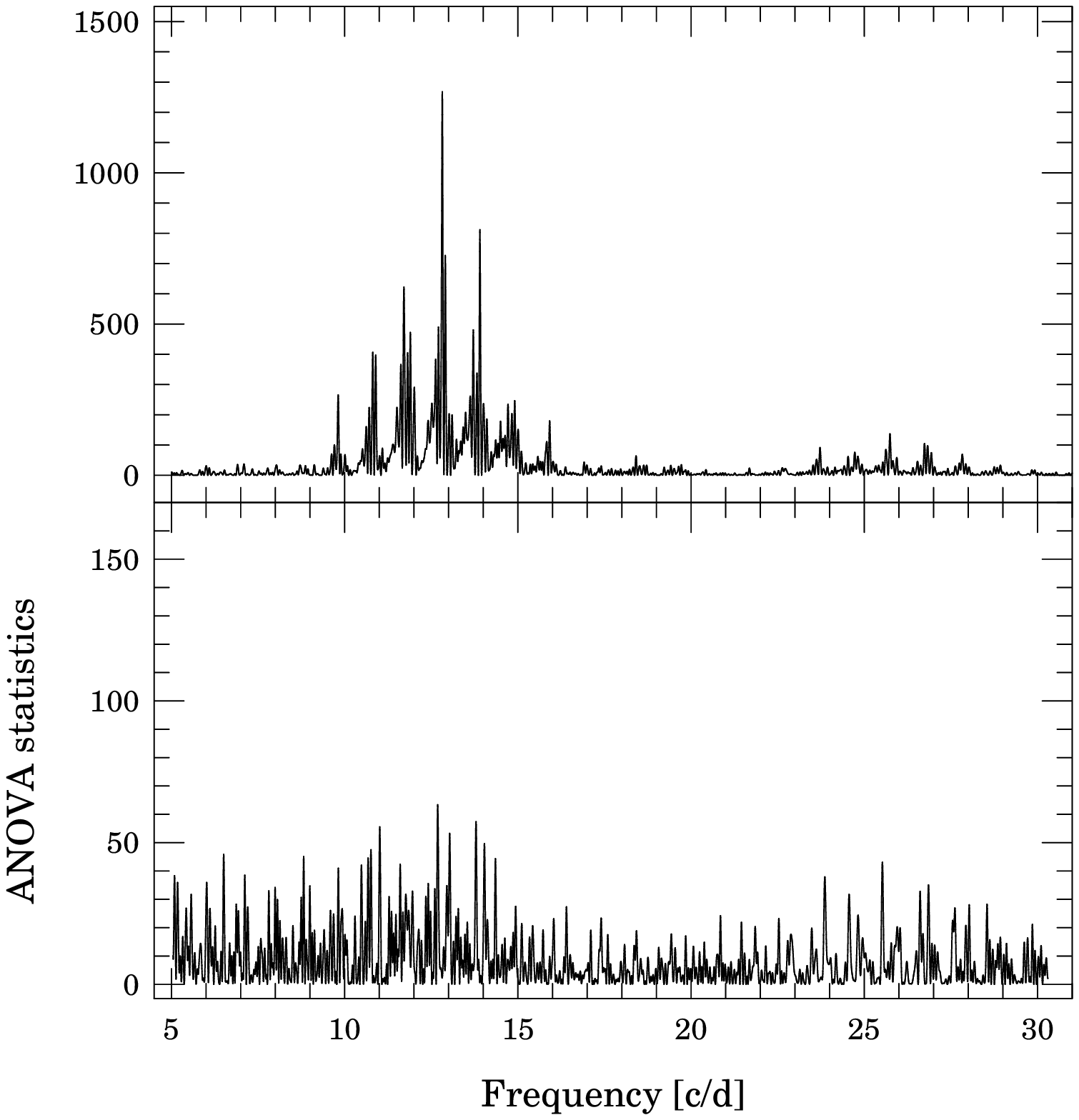}

   \begin{figure}[h]
      \caption{\sf {\sc Anova} power spectrum of the light curve of TT
Boo from its 2004 superoutburst. Upper panel: power spectrum of the
Original light curve. Lower panel: power spectrum of the prewhitened
light curve.
              }
   \end{figure}

It was shown by Olech at al. (2004) that some of SU UMa stars (eg. IX
Dra and ER UMa) also show modulations with their orbital periods during
the entire superoutburst phase. To check if any other periodicity is
present in the superoutburst light curve of TT Boo, we first removed the
decreasing trend from our nightly observations and then grouped them
into blocks containing 2-4 nights. Within the nights in one block the
shape of the superhumps was similar. Then the data from each segment
were fitted with the following sum:

\begin{equation}
rel.~mag = A_0 + \sum^4_{j=1} A^1_j\sin(2j\pi t/P^*_{sh}+\phi^1_j)
\end{equation}

\noindent where $P^*_{sh}$ is the superhump period determined for each
block independently. In the next step, this analytic relation was
removed from the light curve of each block. The whole resulting light
curve was again analyzed using {\sc anova} statistics with two harmonic
Fourier series. The result is shown in lower panel of Fig. 4. This power
spectrum is noisy with the highest peak (not exceeding the $3\sigma$
level) at frequency of $12.818\pm0.007$ c/d, which looks like the
residual of main superhump frequency rather than a real peak. Finally,
we conclude that in the light curve of TT Boo in superoutburst we did
not find any other frequency except that connected with superhumps.

\vspace{14.5cm}

\includegraphics{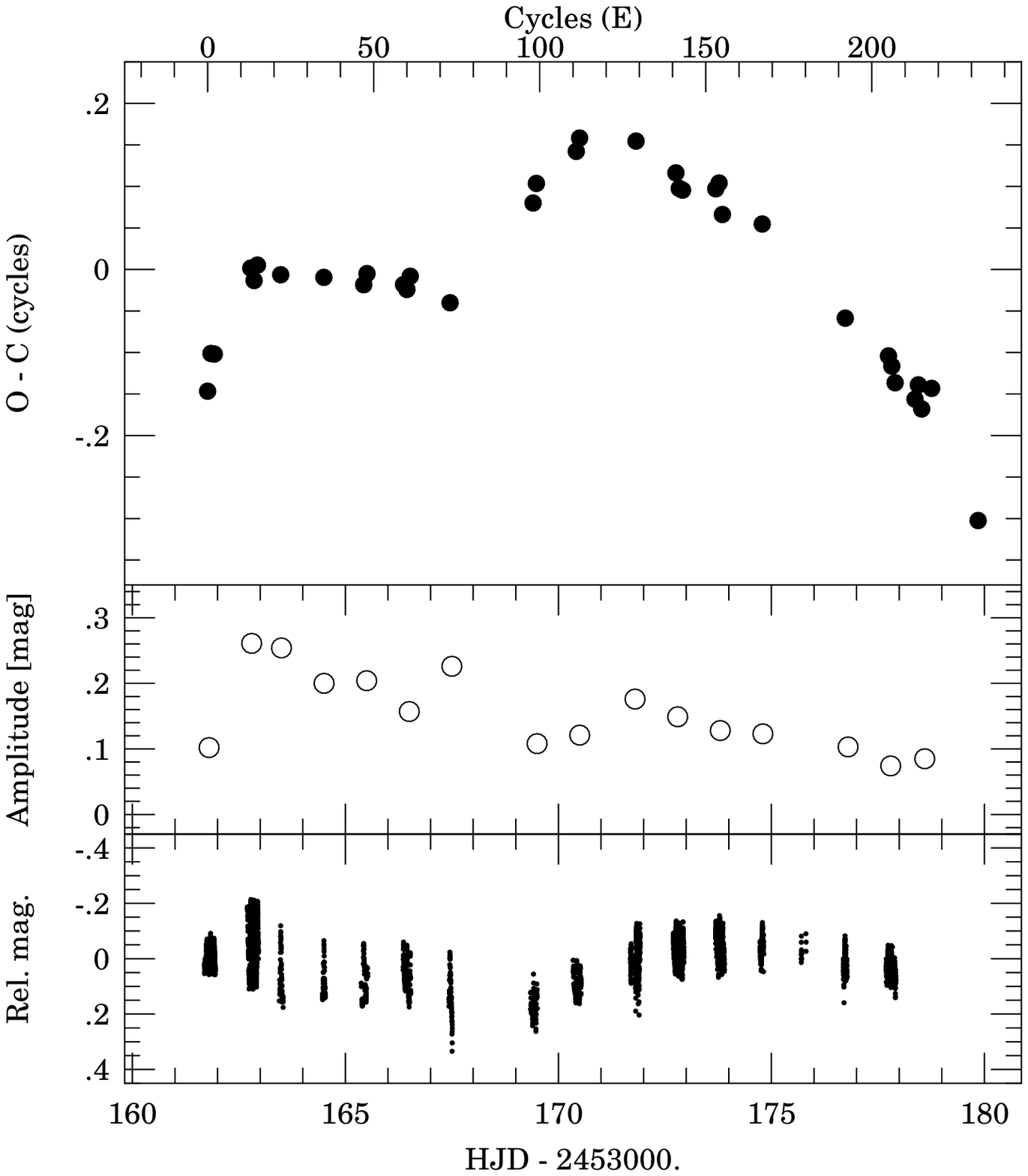}

   \begin{figure}[h]
      \caption{\sf Upper panel: The $O-C$ diagram for superhump maxima
observed in the 2004 superoutburst of TT Boo. Middle panel: Superhump 
amplitude changes during 2004 superoutburst of TT Boo. Lower panel: The 
light curve of TT Boo from its 2004 superoutburst after subtraction of
the mean decline trend.
              }
   \end{figure}

\subsection{The $O-C$ analysis of 2004 data}

To check the stability of the superhump period and to determine its
value we constructed an $O-C$ diagram. We decided to use the timings of
primary maxima, because they were almost always better defined than
minima. In total, we were able to determine 35 times of maxima and they
are listed in Table 3 together with their errors, cycle numbers $E$ and
$O-C$ values.

\begin{table}[!h]
\caption{\sc Times of maxima in the light curve of TT Boo during its
2004 superoutburst.}
\vspace{0.1cm}
\begin{center}
\begin{tabular}{|r|c|c|c|}
\hline
\hline
Cycle & $HJD_{\rm max}-2453000$ & Error & $O-C$\\
number $E$ & & & [cycles] \\
\hline
  0 & 161.7550 & 0.0020 & $-0.1466$\\
  1 & 161.8365 & 0.0020 & $-0.1012$\\
  2 & 161.9144 & 0.0020 & $-0.1019$\\
 13 & 162.7800 & 0.0015 & $+0.0015$\\
 14 & 162.8568 & 0.0015 & $-0.0133$\\
 15 & 162.9362 & 0.0015 & $+0.0052$\\
 22 & 163.4810 & 0.0015 & $-0.0064$\\
 35 & 164.4942 & 0.0025 & $-0.0095$\\
 47 & 165.4290 & 0.0020 & $-0.0184$\\
 48 & 165.5080 & 0.0080 & $-0.0050$\\
 59 & 166.3645 & 0.0015 & $-0.0183$\\
 60 & 166.4420 & 0.0030 & $-0.0241$\\
 61 & 166.5212 & 0.0017 & $-0.0082$\\
 73 & 167.4542 & 0.0020 & $-0.0401$\\
 98 & 169.4125 & 0.0025 & $+0.0799$\\
 99 & 169.4923 & 0.0020 & $+0.1036$\\
111 & 170.4308 & 0.0015 & $+0.1422$\\
112 & 170.5100 & 0.0025 & $+0.1581$\\
129 & 171.8350 & 0.0030 & $+0.1546$\\
141 & 172.7675 & 0.0020 & $+0.1162$\\
142 & 172.8440 & 0.0015 & $+0.0975$\\
143 & 172.9218 & 0.0025 & $+0.0955$\\
153 & 173.7015 & 0.0018 & $+0.0971$\\
154 & 173.7800 & 0.0030 & $+0.1041$\\
155 & 173.8550 & 0.0040 & $+0.0661$\\
167 & 174.7896 & 0.0020 & $+0.0547$\\
192 & 176.7297 & 0.0022 & $-0.0587$\\
205 & 177.7396 & 0.0020 & $-0.1042$\\
206 & 177.8166 & 0.0018 & $-0.1164$\\
207 & 177.8930 & 0.0040 & $-0.1364$\\
213 & 178.3592 & 0.0040 & $-0.1562$\\
214 & 178.4385 & 0.0030 & $-0.1390$\\
215 & 178.5142 & 0.0030 & $-0.1680$\\
218 & 178.7500 & 0.0025 & $-0.1432$\\
232 & 179.8290 & 0.0040 & $-0.3024$\\
\hline
\hline
\end{tabular}
\end{center}
\end{table}

The least squares linear fit to the data from Table 3 gives the
following ephemeris for the maxima:

\begin{equation}
{\rm HJD}_{max} =  2453161.76643(57) + 0.0779575(48) \cdot E
\end{equation}

\noindent indicating that the mean value of the superhump period
is equal to 0.0779575(48) days ($112.259\pm0.007$ min). This is in
good agreement with the value obtained from the power spectrum analysis.
Combining both our period determinations gives a mean value of the
superhump period as $P_{sh} = 0.0779589(47)$ days ($112.261\pm0.007$
min).

The $O-C$ values computed according to the ephemeris (2) are listed in
Table 3 and also shown in the upper panel of Fig. 5.

\subsection{Period change pattern}

Until the mid of 1990's all members of the SU UMa group seemed to show
only negative superhump period derivatives (Warner 1995, Patterson et
al. 1993). This was interpreted as a result of disk shrinkage during the
superoutburst, thus lengthening its precession rate (Lubow 1992). This
picture became more complicated when the first stars with $\dot P>0$
were discovered. Positive period derivatives were observed only in stars
with short superhump periods close to the minimum orbital period for
hydrogen rich secondary (e.g. SW UMa - Semeniuk et al. 1997, WX Cet -
Kato et al. 2001a, HV Vir - Kato et. al 2001b) or for stars below this
boundary (e.g. V485 Cen - Olech 1997, 1RXS J232953.9+062814 - Uemura et
al. 2002).

The diversity of $\dot P$ behavior is well represented in the $\dot P/P$
versus $P_{sh}$ diagram shown for example in Kato et al (2003b) or Olech
et al. (2003). This graph seems to suggest that short period systems are
characterized by positive period derivatives, while these with longer
period by negative period derivatives.

Recently Olech et al (2003) investigated the $O-C$ diagrams for stars
such as KS UMa, ER UMa, V1159 Ori, CY UMa, V1028 Cyg, RZ Sge and SX LMi
and claimed that most (probably almost all) SU UMa stars show decreasing
superhump periods at the beginning and the end of superoutburst but
increasing period in the middle phase.

The $O-C$ diagram obtained for TT Boo seems to confirm this hypothesis.
The superhump period change is quite complex and the rough fitting of
the parabolas to the following cycle intervals: 0--35, 47--112, 98--232
gives the period derivatives of $(-52.3\pm1.3) \times 10^{-5}$, 
$(12.3\pm4.8) \times 10^{-5}$ and $(-6.2\pm0.9) \times 10^{-5}$,
respectively.

It is interesting that the period changes seem to be correlated with
changes in the amplitude of the superhumps and variations of the
brightness of the star. This is clearly visible in Fig. 5 which shows
$O-C$ values, amplitude variations in time and the light curve from
superoutburst after removing the mean long term decline. 

\subsection{Period change, phase shift or both?}

The problem with tracing the period changes using $O-C$ diagrams is that
slight and continuous phase shifts at constant period can mimic true
period changes. In fact, exactly the same $O-C$ plots could be obtained
for synthetic light curves in one case with constant period and time
dependent phase shift and in another case with constant phases and
period variation in time.

Most recently, Pretorius et al (2004) described the results of an 
extensive campaign on the new SU UMa-type variable SDSS J013701.06
-091234.9. In their $O-C$ diagram of superhump maxima they describe the
behavior as consistent with a constant period during first week of
superoutburst and continuous phase shift in later period. However,
detailed inspection of the $O-C$ values from first part of the
superoutburst seems to agree with scenario of decreasing period during
first three days and increasing during next four.

It is interesting, that about a dozen days after maximal brightness the
$O-C$ has a value of 0.5 indicating that at this moment light
modulations could be classified as late superhumps. These late
superhumps have a significantly shorter period than normal superhumps
indicating that except for possible phase shifts, a clear change of the
period occurred.

Do we observe late superhumps in 2004 superoutburst of TT Boo? The
answer seems to be 'no'. The $O-C$ values at the termination of the
superoutburst are around $-0.3$. Even changing the ephemeris (2) to a
longer period, better describing the large amplitude superhumps observed
between cycles 10 and 100, we obtain a phase shift between maxima only
at the level of 0.35 cycle. As a trace of young late superhumps we can
assume the modulations observed on June 22/23 when the star entered into
the final decline phase (see next section for details).

Do we observe period change in late stages of TT Boo superoutburst? In
this case, the answer seems to be 'yes'. First, the superhump period for
nights with cycle numbers larger than 180 is shorter than the superhump
period of the large amplitude superhumps observed at the beginning and
in the middle of superoutburst (just as in case of SDSS J01370106
-091234.9). Second, if we group our observations into two night segments
and calculate the superhump period for each of these segments we obtain
a slightly decreasing pattern. A simple linear fit to the period values
obtained for cycle number 100 and larger gives a slope of $(-6.6\pm2.5)
\times 10^{-5}$, clearly consistent with a parabola fitting the times of
maxima in the same cycle intervals, which gives a value of $(-6.2\pm0.9)
\times 10^{-5}$.

It is now clear that the simple model with a shrinking disk as the cause
of negative superhump period derivatives is no longer valid. A new model
must contend with the following observational facts:

\begin{itemize}

\item complex superhump period change patterns as seen in TT Boo and
other well observed SU UMa stars,

\item extreme values of superhump period derivatives as observed for
example in KK Tel (Kato et al., 2003b) and MN Dra (Nogami et al., 2003).

\item no superhump period changes occur in some SU UMa stars (for example IX Dra - Olech et al., 2004)

\end{itemize}

\section{Late superhumps}

In August 1989 TT Boo was observed during two consecutive nights near
the end of the superoutburst. Combining these data with observations
from June 2004 allowed us to trace the birth of the late superhumps. The
upper panel of Fig. 6 shows observations from June 21/22 and June 22/23 of
2004. When the star was at $V$ magnitude between 14.3 and 14.7, it
showed clear modulations with an amplitude of 0.085 mag and only a weak
trace of secondary humps. The first four arrows point the moments of
maxima displayed in Table 3. The last two arrows are expected times of
maxima computed based on the moment of maximum at $E=218$ and period of
$P=0.0775$ days. One can clearly see that on June 22/23 the amplitude of
the modulations did not change significantly in comparison with the
previous night but the secondary humps became strong enough to show an
amplitude similar to the main maxima.

\vspace{14cm}

\includegraphics{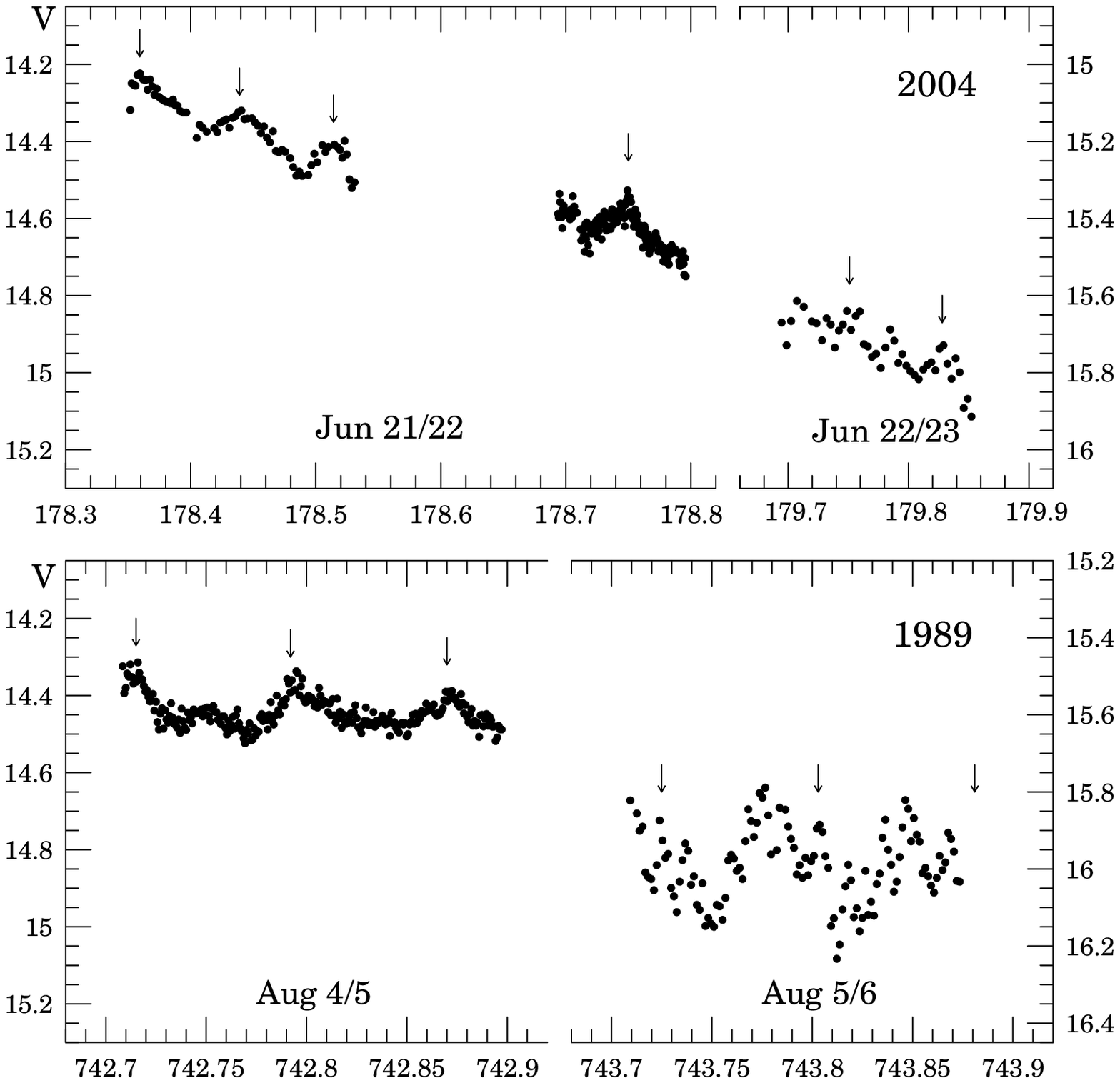}

   \begin{figure}[h]
      \caption{\sf Upper panel: The light curves of TT Boo during the
nights of 2004 June 21/22 and 22/23. Lower panel: The light curves of
TT Boo during the nights of 1989 August 4/5 and 5/6. The arrows marks
observed or expected times of ordinary superhump maxima.
              }
   \end{figure}

What happens when the star fades to $V\approx16$ mag can be seen thanks
to the 1989 data shown in the lower panel of Fig. 6. On August 4/5, when
the star was at $V=14.4$ mag we see the same behavior as in the
corresponding stage of superoutburst from 2004. The arrows mark the
positions of maxima of ordinary superhumps. The secondary humps are
marginally visible. The amplitude of the modulations is about 0.1 mag.
On the next night, when the star was at $V\approx16$ mag, the amplitude
increased to over 0.3 mag. The last three arrows on the lower panel of
Fig. 6 show expected positions of ordinary humps maxima and in this case
they coincide with the secondary maxima. Main maxima are thus shifted by
0.5 in phase in comparison with the previous night.

Summing up, the behavior of TT Boo in superoutbursts observed in 1989
and 2004 suggests that during last stage of plateau phase the period of
superhumps decreases continuously producing a phase shift of about 0.3
in comparison with large amplitude superhumps observed in the beginning
of the superoutburst. In this phase, we still can call the observed
modulations ordinary superhumps, even in the case when the phase shift
caused by a decrease of the period reaches a value of 0.5 as was
observed in SDSS J013701.06-091234.9 (Pretorius et al. 2004). During the
final decline stage, the amplitude of the secondary humps becomes
comparable with the amplitude of main modulations, and within one day
these secondary humps flash into large amplitude late superhumps. Thus
the late superhumps are in fact shifted in phase by 0.5 but in
comparison with ordinary superhumps observed at the end of superoutburst
not with these seen at the beginning. A similar situation was observed
in the well studied dwarf nova VW Hyi (Schoembs \& Vogt 1980, Vogt 1983)
where late superhumps appeared after the rapid decline phase and caused a
beat phenomenon due to the combination with the orbital hump.

\section{Summary}

We described the results of the observations of TT Boo in two
superoutbursts from 1989 and 2004. The main conclusions of our work
are summarized below:

\begin{enumerate}

\item The amplitude of the 2004 June superoutburst was about 5.5 mag
and lasted just over 22 days.

\item The star showed a clear re-brightening around 9th day of the
superoutburst (see Kato et al. 2003a for more detailed discussion
of such a phenomenon).

\item During two observed superoutbursts, we detected clear
superhumps with a period of $P_{sh} = 0.0779589(47)$ days
($112.261\pm0.007$ min) No other periodicity was detected.

\item The superhump period change is quite complex but has a decreasing
period during the first and the end phase of the superoutburst with an
increasing period in the middle stage.

\item Combining the data from two superoutbursts from 1989 and 2004
allowed us to trace the birth of the late superhumps and we conclude
that it is a rather quick process lasting about one day.

\end{enumerate}

\bigskip \noindent {\bf Acknowledgments.} ~We acknowledge generous
allocation of  the Warsaw Observatory 0.6-m telescope time. Data from
AAVSO and VSNET observers is also appreciated. This work was supported
by KBN grant number 1~P03D~006~27 to AO and BST grant to Warsaw
University Observatory. Observations of TT Boo in 1989 were supported by
the grant of the NRSEC of Canada to Dr. S.M.~Rucinski.


{\small

\begin{thebibliography}{}

   \bibitem{be00} Berry, R. and Burnell, J, 2000, The Handbook of
Astronomical Imaging Processing, Willmann-Bell, Inc., Richmond, VA, USA. 
   \bibitem{br89} Bruch A., 1989, A\&A Suppl. Ser., 78, 145
   \bibitem{br94} Bruch A., Engel A., 1994, A\&AS, 104, 79
   \bibitem{ca93} Caldwell J.A.R., Cousins A.W.J., Ahlers C.C.,
           van Wamelen P., Maritz E.J., 1993, SAAO Circ., 15, 1
   \bibitem{hs89} Howell S.B., Szkody P., 1988, PASP, 100, 224
   \bibitem{ka95} Kato T., 1995, IBVS no. 4243
   \bibitem{ka01a} Kato T., Masumoto K., Nogami D., Morikawa K., Kiyota S.,
            2001a, PASJ, 53, 893
   \bibitem{ka01b} Kato T., Sekine T., Hirata R., 2001b, PASJ, 53, 1191
   \bibitem{ka03a} Kato T., Nogami D., Moilanen M., Yamaoka H., 2003a,
      PASJ, 55, 989
   \bibitem{ka03b} Kato T., Santallo R., Bolt G. et al., 2003b, MNRAS,
      339, 861 
   \bibitem{kh98} Kholopov P.N., Samus N.N., Frolov M. et al, 1998,
    Combined General Catalogue of Variable Stars, 4.1 Ed (II/214A)
   \bibitem{lu92} Lubow S.H., 1992, ApJ, 401, 317
   \bibitem{me66} Meinunger L., 1966, Mitt. Ver\"anderl. Sterne, 3, 113
   \bibitem{no03} Nogami D., Uemura M., Ishioka R. et al., 2003,
           A\&A, 404, 1067
   \bibitem{ol97} Olech A., 1997, Acta Astron., 47, 281
   \bibitem{ol03} Olech A., Schwarzenberg-Czerny A., P. K\c{e}dzierski,
           K. Z{\l}oczewski, K. Mularczyk, M. Wi\'sniewski, 2003, Acta
           Astron., 53, 175
   \bibitem{ol04} Olech A., K. Z{\l}oczewski, K. Mularczyk, P.
           K\c{e}dzierski, M. Wi\'sniewski, G. Stachowski, 2004, Acta
           Astron., 54, 57
   \bibitem{pa83} Patterson J., Bond H.E., Grauer A.D., Shafter A.W., 
            Mattei J.A., 1993, PASP, 105, 69
   \bibitem{pr04} Pretorius M.L., Woudt P.A., Warner B., Bolt G.,
           Patterson J., Armstrong E., 2004, MNRAS, in print, 
           astro-ph/0405202
   \bibitem{sc80} Schoembs R., Vogt N., 1980, A\&A, 91, 25
   \bibitem{sc96} Schwarzenberg-Czerny A., 1996, ApJ Letters,
           460, L107
   \bibitem{se97} Semeniuk I., Olech A., Kwast T., Nale\.zyty M., 1997,
           Acta Astron.,  47, 201
   \bibitem{sh23} Shapley H., 1923, Harvard Coll. Obs. Bull. no. 791
   \bibitem{st87} Stetson P.B., 1987, PASP, 99, 191
   \bibitem{sz87} Szkody P., 1987, ApJ Suppl. Ser., 63, 685
   \bibitem{up92}  Udalski A., Pych W., 1992, Acta Astron.,
           42, 28
   \bibitem{ue02} Uemura M., Kato T., Ishioka I. et al., 2002, PASJ, 54, 599
   \bibitem{vo83} Vogt N., 1983, A\&A, 118, 95 
   \bibitem{war95} Warner B., 1995, {\it Cataclysmic Variable Stars}, 
           Cambridge University Press
\end{thebibliography}
}
\end{document}